\documentclass[lettersize,journal]{IEEEtran}
\AtBeginDocument{%
  }

\let\sigproof\proof\let\proof\relax
\let\sigendproof\endproof\let\endproof\relax
\usepackage[utf8]{inputenc} 
\usepackage[T1]{fontenc}
\usepackage{url}
\usepackage{nicematrix}
\usepackage{lipsum}
\usepackage{ifthen}
\usepackage{bbm}
\usepackage{makecell}
\usepackage{listings}
\usepackage{multirow} 
\usepackage{hyperref}
\usepackage{booktabs}
\usepackage{authblk}
\usepackage{nicematrix}
\usepackage{amsmath}[cmex10] 
\usepackage{enumitem}
\usepackage{amsfonts,bm,epsfig,graphicx,latexsym,float}
\usepackage{rotating,setspace,latexsym,epsf,color,subcaption}
\usepackage{comment,caption}
\usepackage{soul}
\usepackage{tcolorbox}

\newcommand{\xhdr}[1]{{\vspace{1pt}\noindent\bfseries #1}.}

\let\proof\sigproof
\let\endproof\sigendproof

\newtheorem{remark}{Remark}

\definecolor{niceblue}{HTML}{007ED6}

\lstdefinestyle{promptstyle}{                           
backgroundcolor=\color{gray!5},       
basicstyle=\ttfamily,    
frame=single,                         
rulecolor=\color{gray},               
breaklines=true,                      
breakindent=0pt,                 
keywordstyle=\color{niceblue}\bfseries,   
commentstyle=\color{green}\itshape,   
stringstyle=\color{orange},           
showstringspaces=false,               
tabsize=4,                            
captionpos=t,                         
morekeywords={Question, Answer, Ground, Truth, Passage, Abstract, Website, Provided} 
}
\DeclareCaptionFormat{listing}{#1#2#3}
\begin{document}

\title{Tele-LLMs: A Series of Specialized Large Language Models for Telecommunications}

\author[$\S$]{Ali Maatouk}
\author[$\dagger$]{Kenny Chirino Ampudia}
\author[$\S$]{Rex Ying}
\author[$\S$]{Leandros Tassiulas}

\affil[$\S$]{Yale University, New Haven, CT 06511, USA}
\affil[$\dagger$]{Amazon Web Services, New York, NY 10001, USA}

\maketitle
\begin{abstract}
The emergence of large language models (LLMs) has significantly impacted various fields, from natural language processing to sectors like medicine and finance. However, despite their rapid proliferation, the applications of LLMs in telecommunications remain limited, often relying on general-purpose models that lack domain-specific specialization. This lack of specialization results in underperformance, particularly when dealing with telecommunications-specific technical terminology and their associated mathematical representations. This paper addresses this gap by first creating and disseminating Tele-Data, a comprehensive dataset of telecommunications material curated from relevant sources, and Tele-Eval, a large-scale question-and-answer dataset tailored to the domain. Through extensive experiments, we explore the most effective training techniques for adapting LLMs to the telecommunications domain, ranging from examining the division of expertise across various telecommunications aspects to employing parameter-efficient techniques. We also investigate how models of different sizes behave during adaptation and analyze the impact of their training data on this behavior. Leveraging these findings, we develop and open-source Tele-LLMs\footnote{The Hugging Face links to both the datasets and Tele-LLMs can be found at \url{https://github.com/Ali-maatouk/Tele-LLMs}

\hspace{4pt}This work was conducted during Kenny's time at Yale University.}, the first series of language models ranging from 1B to 8B parameters, specifically tailored for telecommunications.
Our evaluations demonstrate that these models outperform their general-purpose counterparts on Tele-Eval and telecommunications-related literature tasks while retaining their previously acquired capabilities, thus avoiding the catastrophic forgetting phenomenon.
\end{abstract}

\section{Introduction}
\begin{figure*}[h]
    \centering
    \includegraphics[width = \linewidth]{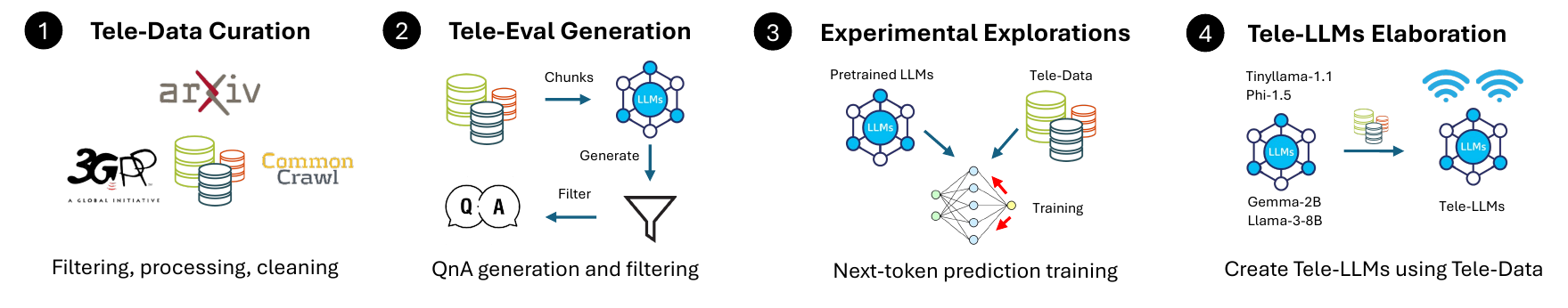}
    \caption{Overall pipeline of the LLM adaptation to telecommunications.}
    \label{fig:overall}
\end{figure*}
Large language models (LLMs) have recently marked a major breakthrough in natural language processing. Since 2022, these deep learning systems have proliferated rapidly, with numerous models released by tech giants, research institutions, and open-source communities \cite{openai2024gpt4technicalreport,gunasekar2023textbooksneed, dubey2024llama3herdmodels}. Trained on vast text corpora, LLMs have demonstrated unprecedented capabilities in understanding context, generating human-like text, and performing reasoning tasks across diverse domains \cite{radford2019language}. These abilities have sparked interest from researchers and industry professionals to adopt and explore their potential applications across a wide variety of fields.

Researchers in the telecommunications and networking domain are no exception, aiming to leverage LLMs' potential across various tasks within the field. These applications include chatbots for engineers \cite{kotaru2023adapting}, network document analysis \cite{bariah2023understanding}, network modeling and development \cite{10599120}, and wireless systems design \cite{du2023power, xu2024large}. The list of tasks being explored for potential LLM support continues to grow \cite{bariah2023large}, with new research emerging rapidly to assess the limits and utility of these models in the telecommunications context.

Thus far, applications of LLMs in the telecommunications domain have primarily involved prompting \cite{du2023power,10599120}, in-context learning \cite{kotaru2023adapting}, and task-specific fine-tuning \cite{zhou2024generativeaiservice6g, bariah2023understanding}, utilizing either proprietary LLMs like OpenAI's GPT \cite{openai2024gpt4technicalreport} or local open-source generic LLMs like Meta's LLaMA series \cite{dubey2024llama3herdmodels}. However, proprietary LLMs raise concerns as they may require sharing prompts and related data with the LLM owners. Additionally, they offer limited maneuverability and adaptability since users do not have access to the model weights. Furthermore, despite their adaptability, general-purpose open-source LLMs lack specialization in telecommunications, as they encompass knowledge from diverse domains such as medicine, history, and law. This lack of specialization hinders their performance; in fact, it is well established that LLM applications in specific domains, from prompting to task-specific fine-tuning, perform better when using domain-specific LLMs rather than general ones \cite{gururangan-etal-2020-dont}. 

Given the above, in a multitude of domains, efforts have been made to create specialized open-source LLMs to push the performance limits of LLMs applications in those fields, such as medicine \cite{labrak2024biomistralcollectionopensourcepretrained}, law \cite{colombo2024saullm7bpioneeringlargelanguage}, and finance \cite{xie2023efficientcontinualpretrainingbuilding}. However, to date, there are no domain-specific open-source models in the telecommunications literature. This stems from the particularities of the telecommunications domain, which includes heterogeneous material such as standards and scholarly work, a heavy reliance on complex equations, varying formatting in datasets, and the generally closed-source nature of work done by companies and institutions thus far with LLMs \cite{kotaru2023adapting,zou2024telecomgptframeworkbuildtelecomspecfic,holm2021bidirectional}. Given the anticipated role of LLMs in telecommunications, this represents a significant gap in the field. 

\xhdr{Proposed Work} This work aims to address the aforementioned gap by introducing the first open-source series of LLMs specialized for the telecommunications field. Our contribution extends beyond open-sourcing the models; we open-source every step of the framework required for this specialization. We also provide, through extensive experiments, key insights into the process of adapting LLMs to the telecommunications domain. This includes the techniques to be used and the different behaviors exhibited by various LLMs during adaptation, based on their sizes and overall tendencies.  

First, we curate a comprehensive set of telecommunications material, referred to as Tele-Data, by leveraging an LLM-based filtering of arXiv papers, standards, Wikipedia articles, and web content to identify relevant sources. We report the precision and recall of our filtering method based on human-labeled data and highlight its high recall rate. The collected material is then subjected to extensive cleaning using regular expressions and LLM-based filtering techniques tailored to the particularities of the telecommunications domain. Specifically, our process addresses formatting differences in standards documents and unifies the equation materials across all types of documents into LaTeX formatting. To quantify the cleanliness of the resulting material, we propose a cross-entropy-based approach and demonstrate that the material is significantly cleaner after this process. 

\sloppy Second, leveraging Tele-Data, we use an LLM-based framework to create an evaluation dataset consisting of 750k question-and-answer (QnA) pairs, referred to as Tele-Eval. This dataset represents the first large-scale open-ended QnA dataset in the telecommunications field. It also includes references to the specific material from which each question was generated, allowing users to select questions from the material they are interested in and facilitating retrieval-augmented generation frameworks  \cite{lewis2021retrievalaugmentedgenerationknowledgeintensivenlp}. To ensure the relevance and quality of the questions generated by the LLM, we apply strict regular expressions and LLM-based filtering to exclude questions of local interest, such as simulation results or locally-defined content.


Third, we examine the viability of using parameter-efficient fine-tuning (PEFT) techniques to inject telecommunications knowledge into these LLMs. Our findings show that, instead, full fine-tuning (FFT) is required. Based on this, we conduct experiments to determine the number of training epochs needed to maximize model performance and to identify when overfitting occurs. We also explore how models of different sizes behave during adaptation and how training dynamics vary depending on the original model’s data. Furthermore, we assess whether splitting the adaptation into multiple specialized models, each focused on different aspects of telecommunications, is more effective than a single, combined model. Our results demonstrate that the latter approach is superior due to the transfer learning that occurs across these aspects.

Finally, building on these findings, our work culminates in adapting and open-sourcing a series of telecommunications-specialized LLMs ranging from 1B to 8B parameters. Our series is based on TinyLlama-1.1B, Phi-1.5B, Gemma-2B, Gemma-2-2B, LLaMA-3.2-1B, LLaMA-3.2-3B, and LLaMA-3-8B, and includes both base models and their instruction-tuned versions for chatbot applications. To evaluate our telecom-adapted models, we first compare them against their original counterparts using Tele-Eval, employing both quantitative analysis—based on a set of proposed metrics—and qualitative assessment. Our results show an average relative improvement of 25\% on Tele-Eval, thus demonstrating specialization for telecom knowledge while preserving the models' original capabilities and avoiding catastrophic forgetting. Next, to further evaluate our telecom-adapted models, we curate downstream tasks related to telecom literature, such as citation prediction and recommendation. Our results show that the telecom-specialized models consistently outperform their original counterparts on these tasks. Moreover, when fine-tuned appropriately, they can rival larger general-purpose models like GPT-4o on these telecom literature applications. The overall adaptation pipeline is provided in Fig. \ref{fig:overall}.

\section{Domain Adaptation}
\subsection{Background}

LLMs are trained using a next-token prediction objective on corpora that encompass a wide variety of domains, such as medicine, history, and more. In this context, a token refers to a subword that represents the most basic unit in the text. This training process allows the LLM to become accustomed to the distribution of tokens in these datasets and develop an understanding of the statistical relationships between them. Consequently, the LLM develops a well-rounded understanding of the multiple domains it has been trained on.

As demonstrated in \cite{yilidz}, if one intends to use an LLM solely for a specific domain, specializing the LLM in that domain is advantageous for any subsequent in-context learning or task-specific fine-tuning. This is because the probability distribution of tokens in the domain of interest can differ significantly from those in other domains internalized by the LLM. By adapting to this distribution, the LLM's knowledge becomes more targeted to the domain, leading to transfer learning for subsequent usage stages.

The primary approach for domain adaptation is continual pretraining. This method involves further training the LLM on domain-specific corpora, allowing it to adapt its parameters to the field of interest \cite{gururangan-etal-2020-dont}. Continual pretraining has become the standard for LLM domain adaptation, as evidenced by its application in various fields such as medicine \cite{labrak2024biomistralcollectionopensourcepretrained}, law \cite{colombo2024saullm7bpioneeringlargelanguage}, and finance \cite{xie2023efficientcontinualpretrainingbuilding}. In the following, we will explore the details of continual pretraining, with a particular emphasis on its application in the telecommunications domain.

\begin{remark}
Adapting a pretrained LLM to a specific domain is typically preferred over training an LLM from scratch on domain-specific data. This approach leverages capabilities learned from other domains (e.g., English language proficiency, coding) and transfers them to the new domain, reducing the need for extensive data and time to learn basic skills.
\end{remark}

\subsection{Continual Pretraining}
Given the scale of the training datasets of current LLMs (e.g., 15T tokens for LLama-3 \cite{dubey2024llama3herdmodels}), it is fair to assume that the language models available in the literature have been trained on all available public-source data. Therefore, these models have likely already encountered most of the publicly available telecommunications-related data. Nevertheless, the goal of continual pretraining is to adapt the LLM's knowledge to this specific domain by re-exposing the model to these data. Specifically, consider an LLM trained on a corpus of tokens $\mathcal{D}$ drawn from a distribution $\mu$. This LLM is trained over the entire corpus $\mathcal{D}$ to minimize the cross-entropy loss function for tokens $\boldsymbol{x} = (x_1, \ldots, x_T)$ drawn from this corpus:
\begin{equation}
 \min_{\boldsymbol{W} \in \mathbb{R}^n} \mathbb{E}_{\boldsymbol{x} \sim \mu} \left[ \mathcal{L}_{\text{CE}}(\boldsymbol{x}; \boldsymbol{W}) \right] = -\mathbb{E}_{\boldsymbol{x} \sim \mu} \left[ \sum_{t=1}^{T} \log P(x_t \mid \boldsymbol{x}_{1:t-1}; \boldsymbol{W}) \right]
\end{equation}
where $T$ is the context length, $\boldsymbol{x}_{1:t-1}$ is equal to $(x_1,\ldots,x_{t-1})$, $\boldsymbol{W}$ is the LLM's parameters, and $n$ is the total number of parameters. On the other hand, let us consider a set of telecom-related data $\mathcal{D}_{\textnormal{Tele-Data}}$. The samples of tokens $\boldsymbol{x} = (x_1, \ldots, x_T)$ drawn from $\mathcal{D}_{\textnormal{Tele-Data}}$ originate from a different distribution, denoted by $\sigma$. Continual pretraining involves initializing the LLM to its existing parameters, denoted by $\boldsymbol{W}_0$, and further reducing the cross-entropy loss, this time using samples of tokens $\boldsymbol{x}$ drawn from $\mathcal{D}_{\textnormal{Tele-Data}}$:
\begin{equation}
 \min_{\boldsymbol{W} \in \mathbb{R}^n} -\mathbb{E}_{\boldsymbol{x} \sim \sigma} \left[ \sum_{t=1}^{T} \log P(x_t \mid \boldsymbol{x}_{1:t-1}; \boldsymbol{W},\boldsymbol{W}_0) \right].
 \label{eq:continual}
 \end{equation}
\begin{sloppypar}
\noindent By doing so, the model parameters shift closer to the new distribution, allowing the LLM to be better calibrated to telecommunications-specific data.  
\end{sloppypar}
\subsection{Catastrophic Forgetting} Ideally, one would hope that such domain adaptation comes at no penalty. However, generally, a penalty is incurred. By shifting the focus from the general token distribution $\mu$ of the training dataset $\mathcal{D}$ to a new distribution $\sigma$, the model risks forgetting previously acquired knowledge. This risk depends on the size of the model, the size of the domain-specific dataset, and how different $\sigma$ is from $\mu$ \cite{luo2024empiricalstudycatastrophicforgetting}. Addressing this issue is crucial; otherwise, one may end up with a telecom-knowledge proficient LLM but lose reasoning capabilities, coding abilities, and general English language understanding in the process. Later in this paper, specifically in Section \ref{sec:trainingsection}, we will detail how we addressed this issue in our continual pretraining framework. With this in mind, the next step in our framework involves curating a comprehensive telecommunications corpus, which we will refer to as Tele-Data. 
\section{Tele-Data}
\label{sec:telecomdata}
\begin{table}[t!]
\centering
\begin{tabular}[t]{lccc}
\toprule
& Items & Size & Tokens\\
\midrule
Arxiv & 90k &	4 GBs &	1.08B \\
Standards & 2.8k  & 334 MBs &	86.45M \\
Wikipedia & 19.5k  & 123 MBs &	26.44M \\
Web & 740k &	6.8 GBs	 &	1.55B  \\
\bottomrule
\end{tabular}
\caption{Tele-Data division across categories.}
\label{table:detailed}
\end{table}
\begin{figure*}[ht]
\centering
\begin{minipage}{0.32\textwidth}
\vspace{45pt}
  \centering
  \vfill
  \begin{table}[H]
  \vfill
    \centering
    \vspace{-15pt}
    \begin{tabular}{lccc}
      \toprule
      & Precision & Recall & F1-Score \\
      \midrule
      Arxiv & 0.666 & 0.956 & 0.785 \\
      Wikipedia & 0.632 & 0.897 & 0.741 \\
      Webpages & 0.455 & 1 & 0.625 \\
      \bottomrule
    \end{tabular}
    \vspace{27pt}
    \captionof{table}{Datasets filtering statistics.}
\label{table:precisionvsrecall}
  \end{table}
\end{minipage}%
\hfill
\begin{minipage}{0.32\textwidth}
  \centering
  \begin{figure}[H]
    \centering
    \includegraphics[width=.96\linewidth]{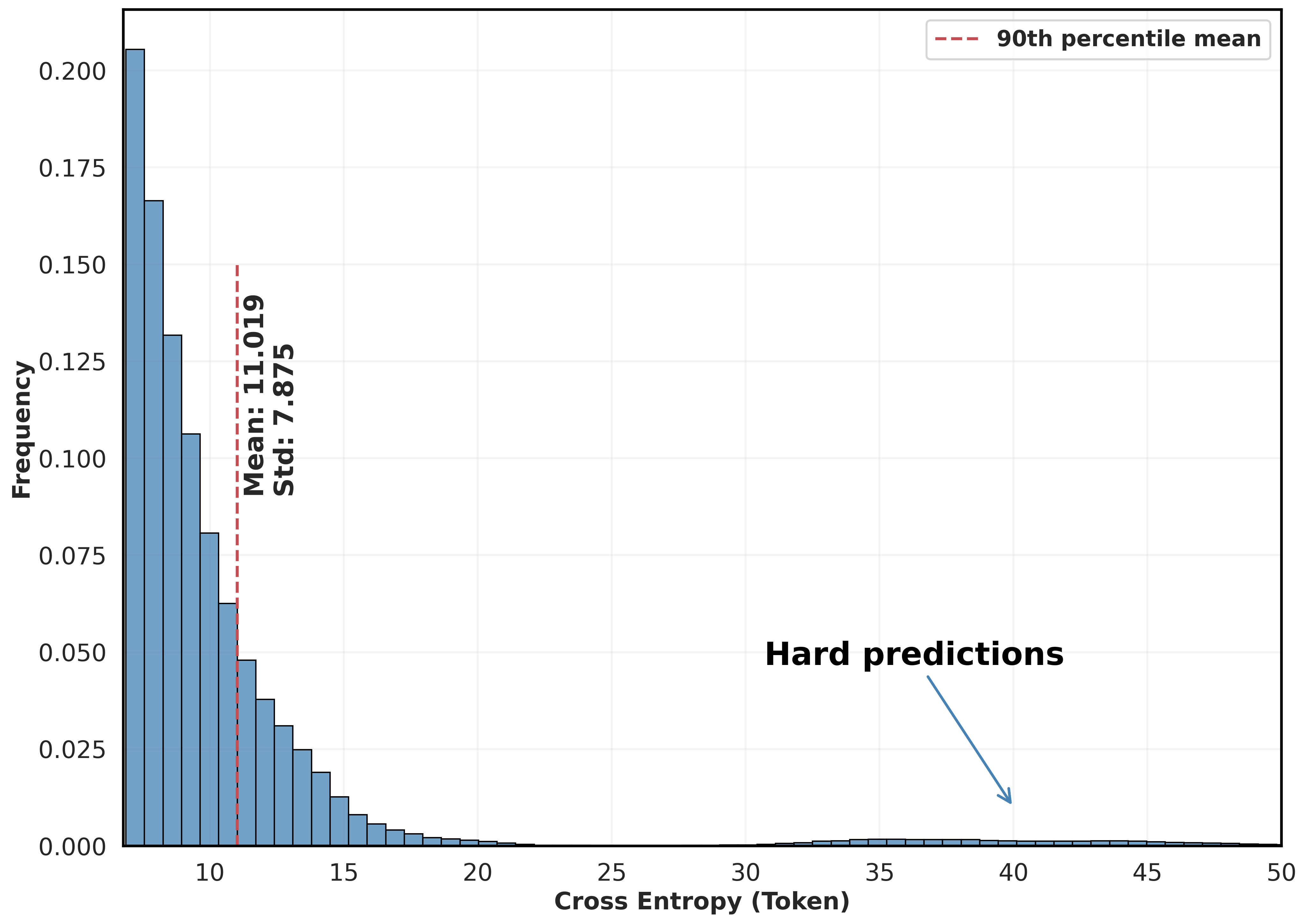}
    \caption{Raw cross-entropy loss.}
    \label{figure:raw}
  \end{figure}
\end{minipage}
\begin{minipage}{0.32\textwidth}
  \centering
  \begin{figure}[H]
    \centering
    \includegraphics[width=.96\linewidth]{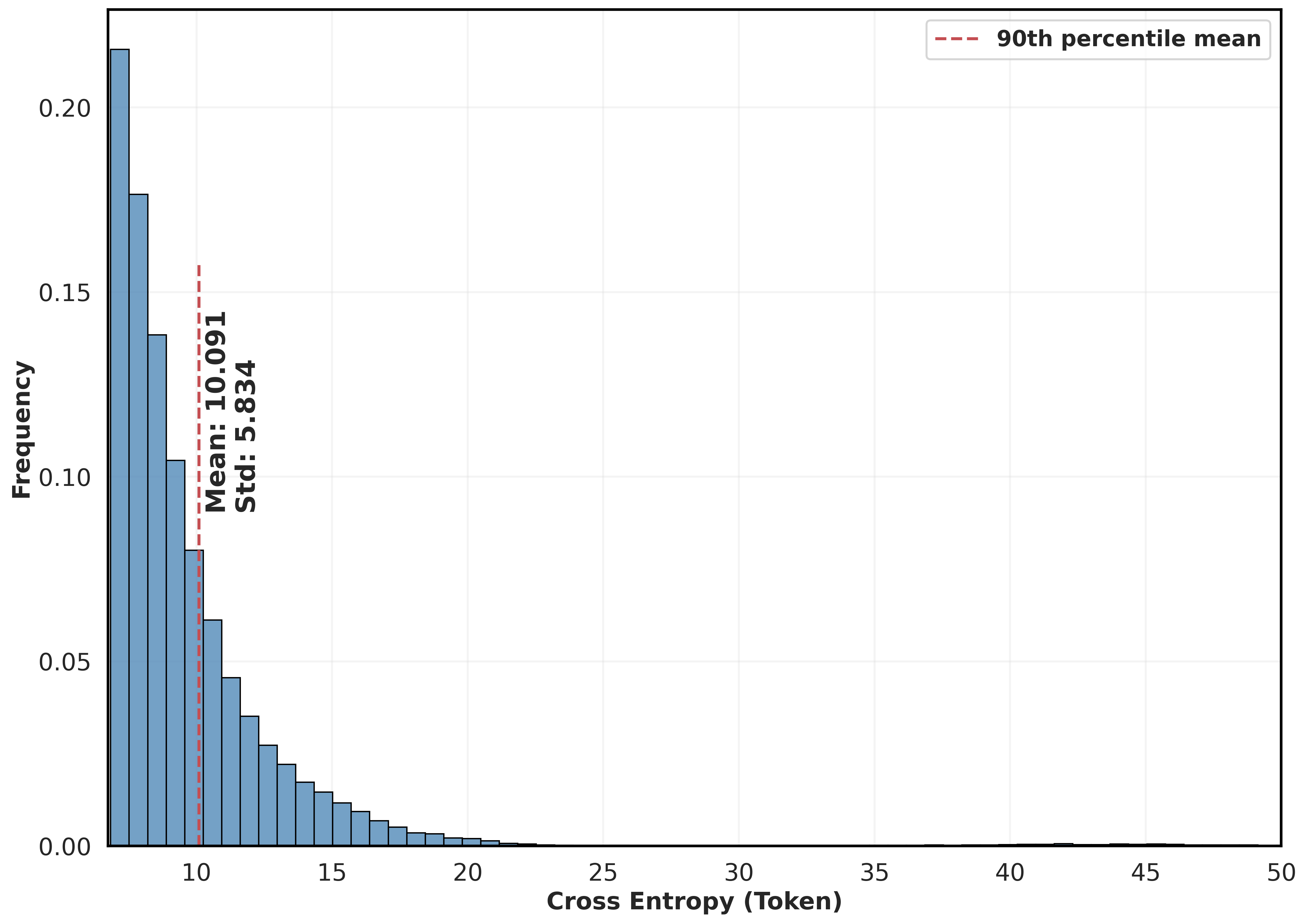}
    \caption{Cleaned cross-entropy loss.}
    \label{figure:clean}
  \end{figure}
\end{minipage}%
\hfill
\end{figure*}
Our collection of telecommunications material comprises four main sources: (1) scientific papers from arXiv, (2) 3GPP standards, (3) Wikipedia articles related to telecommunications, and (4) telecommunications-related websites extracted from Common Crawl dumps. This diverse range of sources ensures broad coverage of telecommunications knowledge and is essential for enabling the transfer of expertise from one aspect of telecommunications covered in one source to another, as will be shown later in Section \ref{sec:divisionexpertise}. The detailed composition of our dataset is reported in Table \ref{table:detailed}, with LLama-3 tokenizer used to provide token counts as an example.

\subsection{Arxiv} 
\xhdr{Curation} One of the largest sources of open-access research on telecommunications consists of preprints submitted by authors to the arXiv repository. As of March 2024, the combined arXiv snapshot for both computer science and electrical engineering categories contains approximately 610k papers. However, given the overlap between these categories and the inclusion of topics beyond telecommunications 
, targeted filtering is necessary to identify the relevant material. To achieve this, we use a language model-based filtering approach. Specifically, we leverage the Mixtral 8x-7B-Instruct\footnote{\url{https://huggingface.co/mistralai/Mixtral-8x7B-Instruct-v0.1}} model, providing it with the abstract of each paper to determine if it is related to the telecommunications and networking domain. The model is prompted\footnote{All the prompts utilized in our framework are reported in Appendix \ref{appendix:prompts}}  to provide a Yes or No answer regarding whether the paper is relevant. We then utilize the logits of the `Yes' and `No' tokens to classify whether or not a paper is related to telecommunications.


To assess the quality of our filtering process, we randomly sampled 500 arXiv papers and manually annotated them to determine their relevance to the telecommunications and networking domain. To ensure a balance between relevant and irrelevant examples, we drew an equal number of samples from both the computer science and electrical engineering categories given that the latter is more likely to contain telecommunications material, while the former is more likely to include unrelated papers. We then evaluated both the precision and recall of the filtering process. As shown in Table \ref{table:precisionvsrecall}, the process demonstrates high recall and moderate precision. High recall is particularly important, as it indicates that our filtering process captures virtually all relevant telecommunications material. The moderate precision can be attributed to the inclusion of papers from related domains. For instance, some papers investigating cloud computing topics were included. Since these additional domains are not orthogonal to telecommunications, they can provide transferable knowledge that may enhance the language model's adaptation to the telecommunications domain. All in all, this demonstrates the effectiveness of our filtering process.

\xhdr{Cleaning} Following curation, we implement a comprehensive cleaning process for the papers, which includes: (1) removing comments, (2) flattening LaTeX sources, (3) substituting user-defined macros with standard LaTeX commands, (4) eliminating LaTeX native commands, and (5) standardizing citation formats and removing markup changes. We also remove figures and tables to focus on inline text and equations. Details on this procedure are provided in Appendix \ref{appendix:cleaning}. To assess the effectiveness of our cleaning process, we randomly selected 500 arXiv papers and derived a cross-entropy-based approach to evaluate cleanliness as detailed below. Let $\mathcal{X} = (\boldsymbol{x}^1, \boldsymbol{x}^2,\ldots, \boldsymbol{x}^N)$ and $\mathcal{Y} = (\boldsymbol{y}^1, \boldsymbol{y}^2,\ldots, \boldsymbol{y}^M)$ denote the non-overlapping chunks of $T$ tokens from the raw and cleaned datasets, respectively, where $N$ and $M$ are the total number of chunks. Next, let us define the cross-entropy loss for any token index $t \in \{1,\ldots, T\}$ in any chunk of either the raw or cleaned data using GPT-2 \cite{radford2019language} as follows
\begin{align}
&\mathcal{L}^{\text{raw}}_{\text{CE}}(x_t^i | \boldsymbol{x}_{1:t-1}^i) = -\log P(x_t^i | x_{1:t-1}^i; \boldsymbol{W}_{\text{GPT-2}}), \quad i=1,\ldots, N,\nonumber\\\nonumber\\&\mathcal{L}^{\text{clean}}_{\text{CE}}(y_t^i | \boldsymbol{y}_{1:t-1}^i) = -\log P(y_t^i | y_{1:t-1}^i; \boldsymbol{W}_{\text{GPT-2}}), \quad i=1,\ldots, M.
\end{align}
GPT-2 was chosen because it was primarily trained on web content, excluding arXiv scientific papers. Therefore, the cross-entropy here serves as a proxy for how closely the text resembles generic online content rather than the more complex arXiv papers, thus providing an indication of the dataset's cleanliness. With this in mind, we set $T$ to 1024, in line with GPT-2's training data \cite{radford2019language}, and define the total set of cross-entropy losses for the raw and cleaned datasets as
\begin{align}
&\mathcal{R} = \bigcup_{i=1}^N \{\mathcal{L}_{\text{CE}}(x_t^i | \boldsymbol{x}_{1:t-1}^i) : t \in [1, 1024]\},\nonumber\\&\mathcal{C} = \bigcup_{i=1}^M \{\mathcal{L}_{\text{CE}}(y_t^i | \boldsymbol{y}_{1:t-1}^i) : t \in [1, 1024]\},
\end{align}
respectively. We then focus on the hardest 10\% of the tokens to evaluate the differences between the two sets $\mathcal{R}$ and $\mathcal{C}$. To that end, let $\tau_1 = Q_{0.9}(\mathcal{R})$ and $\tau_2 = Q_{0.9}(\mathcal{C})$ represent their 90th percentiles. By comparing the distributions of $\mathcal{R}_{\text{high}} = \{\mathcal{L}_{\text{CE}} \in \mathcal{R} : \mathcal{L}_{\text{CE}} \geq \tau_1\}$ and $\mathcal{C}_{\text{high}} = \{\mathcal{L}_{\text{CE}} \in \mathcal{C} : \mathcal{L}_{\text{CE}} \geq \tau_2\}$ in Figures \ref{figure:raw} and \ref{figure:clean}, we observe that both the mean cross-entropy loss and its standard deviation within this percentile are lower in the cleaned dataset compared to the raw, uncleaned papers. This demonstrates the improved cleanliness of our dataset.
\subsection{Standards}
\xhdr{Curation} Standards play a pivotal role in telecommunications as they ensure interoperability among technologies from various vendors. These standards are established and maintained by recognized bodies such as 3GPP, IEEE, and ITU. Due to their open-source nature, we focus on incorporating 3GPP documents into our dataset. To do so, using the 3GPP FTP portal\footnote{\url{https://www.3gpp.org/ftp/}}, we downloaded the latest specifications for each standard of each series, resulting in a dataset of approximately 2.8k documents.

\xhdr{Cleaning} Following the curation process, we clean and process the standards files through several steps. We begin by removing non-essential sections such as related works and appendices, and eliminating figures and tables to focus on inline text and equations, similar to our approach with arXiv papers. A key caveat is that equations in .doc files are formatted in XML, unlike the LaTeX format used in arXiv papers. To address this, we first convert all .doc files to .docx format and then utilize docx2tex\footnote{\url{https://github.com/transpect/docx2tex}} to transform the standards into LaTeX format. This standardization improves the training process by ensuring consistency of equations across all document types. Finally, we apply the same cleaning pipeline used for arXiv papers to the converted standards LaTex files to ensure a uniform level of cleanliness and coherence across our entire dataset.
\subsection{Wikipedia} 
\sloppy
Another source of telecommunications material is the Wikipedia corpus, specifically articles related to the telecommunications domain and its associated technical content. To curate this dataset, we utilize the English subset of the Wikipedia dataset\footnote{\url{https://huggingface.co/datasets/wikimedia/wikipedia}}, which contains 6.4 million samples. Given the size of this dataset, applying a pure LLM-based classification would be computationally expensive. Instead, we employ a two-step process:
\begin{enumerate}[left=0pt]
    \item \textbf{Keyword Filtering:} \sloppy We define a set of 100 telecom-related keywords, including terms such as telecommunications, base station, Wi-Fi, and 5G. Articles containing any of these keywords are flagged for the next step. This process significantly reduces the number of articles from 6.4M to approximately 70k.
    \item \textbf{LLM-based Content Evaluation:} In this second step, we apply an LLM-based filtering process to the flagged articles. Particularly, the first 10,000 characters of each article are provided to the Mixtral 8x-7B-Instruct model, and the LLM is prompted to provide a Yes or No answer regarding the article's relevance and the presence of technical content related to telecommunications. The reason behind this is to exclude articles that discuss non-technical aspects, such as a telecom operator's history.
\end{enumerate}
Through this two-step process, we curate a dataset of 19.5k technically relevant telecommunications articles from Wikipedia. Similarly, we evaluated our LLM-based filtering by sampling 500 flagged articles and manually annotating them. We then assessed the precision and recall of our filtering process. 
The results in Table \ref{table:precisionvsrecall} showcase the high recall of the process, showcasing its ability to identify relevant technical content in telecommunications. The moderate precision, on the other hand, stems from the challenge of determining the appropriate level of technicality required for an article, resulting in false positives within the dataset.
\subsection{Websites} 
The last source of telecommunications material we consider is the Common Crawl dataset, which comprises web archives from across the internet. To avoid issues with duplicates, non-English content, and potential profanity found in raw dumps, we utilized the refined web dataset \cite{refinedweb}. This curated version of Common Crawl contains approximately 1 billion rows across 2.8 terabytes of data. To further eliminate duplicates, we filtered out Wikipedia articles from the dataset. Next, to extract telecommunications-related content from this refined dataset, we employed the same two-step process used for Wikipedia articles. Additionally, we incorporated content from well-known telecommunications blogs, such as ShareTechNote, to enhance the dataset's relevance. The resulting collection consists of content from 740k website links, providing a comprehensive representation of telecommunications information available on the web.

To evaluate the quality of the LLM-based filtering process, we sampled and manually annotated 500 websites containing telecommunications keywords. We then assessed the precision and recall of our filtering process. We observed perfect recall, as the incorporated websites included patents, technical blogs, and forum discussions about telecommunications. However, precision was lower due to the difficulty in discerning the technical content. As a result, websites with product technical specifications, marketing sites for VPN services containing technical details, and product reviews were also included, leading to false positives in the dataset.
\subsection{Dataset Format}
Tele-Data is structured as a JSONL (JSON Lines) file, where each line represents a JSON object with five distinct fields:
\begin{itemize}[left=0pt]
    \item \textbf{ID:} A unique identifier for each data entry, combining the data category and a number. For example, `wiki\_132' refers to the 132th item of the Wikipedia data points.
    \item \textbf{Category:} A string indicating the source of the data: wiki, standard, arxiv, or web.
    \item \textbf{Content:} A string containing the main text of the material.
    \item \textbf{Metadata:} A JSON object containing various information relevant to each specific element, with the structure varying depending on the category. For example, for standards, the JSON object includes the 3GPP series number, release, and standard file name, while for arXiv papers, it contains the arXiv ID, title, and abstract.
    \end{itemize}
Examples of the dataset can be found in Appendix \ref{appendix:teledata}.
\section{Evaluation Dataset}
After preparing Tele-Data, the next step involves creating an evaluation dataset to test the resulting domain-adapted models' telecommunications knowledge. Currently, only one such dataset exists: TeleQnA \cite{maatouk2023teleqnabenchmarkdatasetassess}. TeleQnA is a multiple-choice question (MCQ) dataset drawn from standards and research papers, generated with human involvement. Although an MCQ dataset simplifies evaluation in terms of accuracy, it remains limiting because LLMs are not robust MCQ selectors due to their inherent `selection bias,' a bias prevalent in virtually all LLMs \cite{zheng2024large, griot2024multiplechoicequestionslarge}.

Given the above, to examine the telecommunications proficiency of LLMs, we take another approach by creating a dataset of open-ended telecommunications questions. Open-ended questions assess a model's ability to elaborate on its own to answer questions about specific telecommunications concepts. Given the large number of concepts to be covered and the specialized nature of telecommunications knowledge, large-scale open-ended questions are needed, making a purely human-based approach infeasible. 

To overcome this challenge, we adopt an LLM-based approach to create our evaluation dataset, which we refer to as Tele-Eval. Specifically, we gather material from papers, standards, and wiki articles within the collected Tele-Data. We focus on these categories due to their relative cleanliness compared to the Common Crawl data. We then segment the material into 20,000-character chunks. Using Mixtral 8x7B-Instruct, we prompt the LLM to generate five questions for each segment in a three-shot manner, providing examples similar to those in Trivia QA \cite{joshi-etal-2017-triviaqa}, but tailored to the telecommunications domain. The generated questions then undergo extensive regular expression-based filtering. Particularly, we implement filters to remove questions containing expressions such as `in this context,' `in this paper,' `as highlighted in,' and references to equations, figures, tables, and other local elements, along with 100 other similar filters. Our aim is to achieve the lowest possible retention rate to ensure the highest quality in the automatically generated dataset.

The next step in our process involves feeding the QnA pairs to another instance of the Mixtral LLM. This instance is prompted to determine if the provided QnA pair can be answered without access to the source material (i.e., eliminate local context dependency). We retain only those QnA pairs that pass both the regex-based filtering and this LLM-based filtering. As a result, a retention rate of 1\% was achieved, and the dataset consists of 750k QnA pairs. It is important to note that given this scale, even if some questions are not perfectly answerable by an LLM, the dataset still serves its purpose as an effective evaluator of telecommunications knowledge.

\begin{remark} One important aspect of our dataset is that each generated QnA pair includes the ID of the specific telecommunications material upon which it was based. This allows users to select questions from the material they are interested in and facilitating retrieval-augmented generation frameworks  \cite{lewis2021retrievalaugmentedgenerationknowledgeintensivenlp}. For example, if someone is interested in a specific topic (e.g., source and channel coding), they can identify the relevant content in the telecommunications dataset, obtain their IDs, and then use the portion of Tele-Eval linked to those IDs. 
\end{remark}
\noindent We provide below a couple of examples of the dataset to illustrate the style and format of the data:\\ 

\noindent\textbf{Statement:}  Under what circumstances should the UE insert a public GRUU value in the Contact header field?\\
\textbf{Answer:} The UE should insert the public GRUU value in the Contact header field if a public GRUU value has been saved associated with the public user identity from the P-Called-Party-ID header field, and the UE does not indicate privacy of the P-Asserted-Identity. \\
\textbf{ID:} \color{niceblue} \textbf{standard\_1309}   \color{black} \\

\noindent\textbf{Statement:}  What is the difference between unassisted capacity and entanglement assisted capacity in the context of quantum channels?\\
\textbf{Answer:} Unassisted capacity refers to the maximum rate of transmitting classical or quantum information through a quantum channel without the use of entanglement, while entanglement assisted capacity refers to the maximum rate with the assistance of entanglement.\\
\textbf{ID:} \color{niceblue} \textbf{arxiv\_36721}   \color{black} \\

\section{Evaluation Metrics}
\label{sec:evaluationmetrics}
One of the key challenges after continual pretraining is effectively evaluating the resulting domain-adapted model. As seen in the previous section, Tele-Eval consists of open-ended question-answer pairs. Evaluating open-ended responses is more complex compared to MCQ datasets. For example, traditional evaluation metrics such as ROUGE \cite{lin-2004-rouge} and BLEU \cite{10.3115/1073083.1073135}, commonly used for summarization and translation tasks, measure lexical overlap between the model's output and the reference answers. However, they fail to capture the semantic similarity and correctness of responses when the model uses alternative wording or lexicon to convey the same meaning as the reference answers. This challenge is especially pronounced when evaluating equations generated by the models, as is often required in the telecommunications domain, since these metrics struggle to capture such nuances. To address this, we adopted three evaluation metrics: Answer perplexity, SemScore \cite{aynetdinov2024semscoreautomatedevaluationinstructiontuned}, and LLM-Eval \cite{zheng2023judgingllmasajudgemtbenchchatbot}. In Section \ref{sec:trainingsection}, we will showcase the pros and cons of each evaluation metric and conclude that LLM-Eval is the most robust comparative tool for the telecommunications domain. Below, we provide details on each of these metrics.
\subsection{Answer Perplexity}
We define answer perplexity as the perplexity of the model with respect to the ground truth answer, conditioned on the question. Specifically, let us consider $N$ samples of Tele-Eval, where $\boldsymbol{x}^i=(x^i_1,\ldots,x^i_T)$ represents a concatenation of both the question and the ground truth answer of the i-th pair.
Assume that at index $k_i$, the ground truth answer begins. With that in mind, we define the answer perplexity as:
\begin{equation}
\begin{split}
\text{Ans-PPL}(\boldsymbol{x}) = \exp\bigg(&-\frac{1}{N}\sum_{i=1}^{N}\sum_{j=k_i}^{T_i}\frac{1}{T_i-k_i+1} \\
&\log P(x^i_j \mid \boldsymbol{x}^i_{1:j-1}; \boldsymbol{W})\bigg),
\end{split}
\end{equation}
where $T_i$ is the number of tokens in the i-th pair, and $\boldsymbol{W}$ represents the model weights. This metric can be seen as how surprised the model is by the answer, given the question. Intuitively, a model that is well-versed in telecommunications should be less surprised by the answer (hence, have a lower perplexity) compared to a non-specialized model.
\subsection{SemScore}
Another metric we use to evaluate the correctness of a model's output relative to the ground truth answer is semantic similarity. This is achieved by leveraging sentence transformer models \cite{reimers-2020-multilingual-sentence-bert}, such as the all-mpnet-base model\footnote{\url{https://sbert.net/docs/sentence_transformer/pretrained_models.html}} from the Sentence Transformers family. These encoder-only BERT models are trained using contrastive loss to encode pairs of sequences into a high-dimensional space, such that the cosine similarity between embeddings of similar pairs is high (closer to 1), while it is lower for dissimilar ones (closer to -1). With this in mind, we define the SemScore \cite{aynetdinov2024semscoreautomatedevaluationinstructiontuned} as
\begin{equation}
\textnormal{SemScore}(\boldsymbol{x},\boldsymbol{y}) = \cos(\textnormal{embed}(\boldsymbol{x}),\textnormal{embed}(\boldsymbol{y})),
\end{equation}
where $\textnormal{embed}(\cdot)$ refers to the BERT embedding model, and $\boldsymbol{x}$ and $\boldsymbol{y}$ refer to the ground-truth answer and the model's output, respectively. Thus, this metric allows us to judge how closely the model's output aligns with the correct answer.

\subsection{LLM-Eval}
\begin{figure*}[t]
    \centering
    \includegraphics[width=0.99\textwidth]{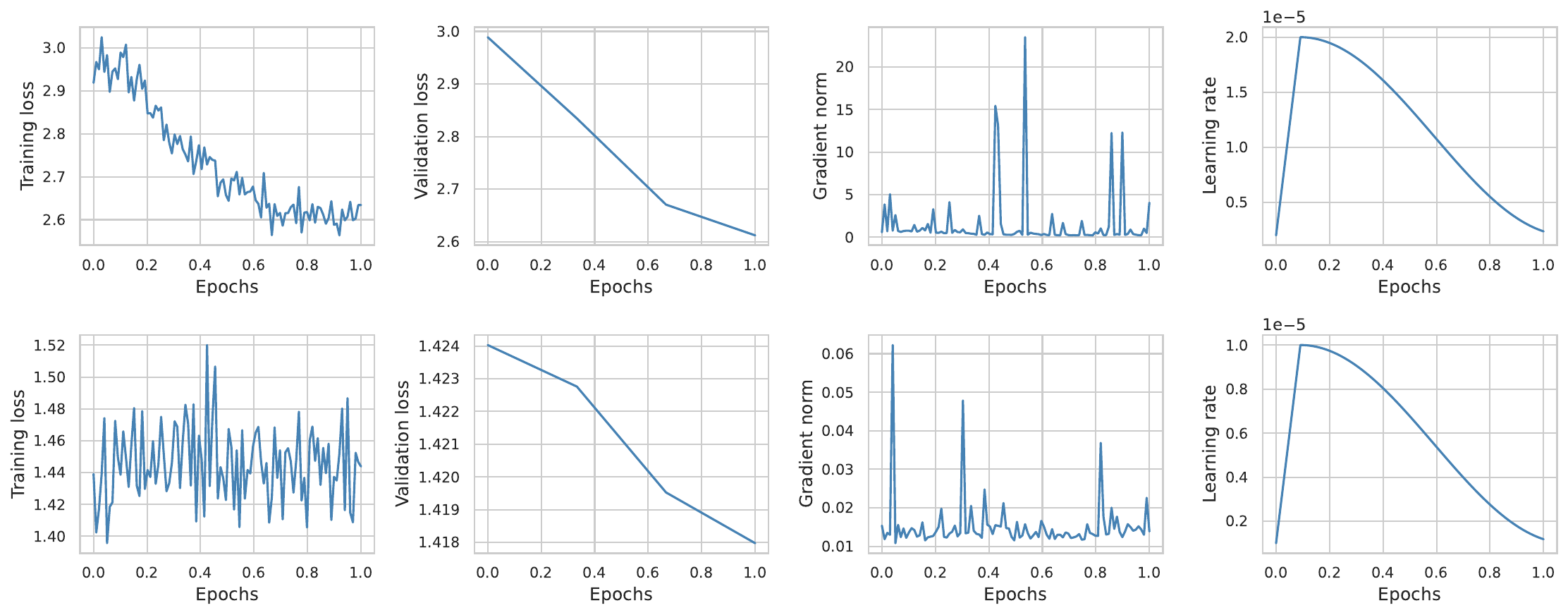}
    \caption{Training metrics for Gemma-2B (top) and Llama-3-8B (bottom) models using LoRa.}
    \label{fig:peftvsfft}
\end{figure*}
The last metric in our framework involves using an LLM as a judge to assess the correctness of a model's output compared to the ground truth answer \cite{zheng2023judgingllmasajudgemtbenchchatbot}. Given the extensive knowledge LLMs have acquired through their training, we can leverage them as assistants to evaluate the quality and accuracy of model outputs. Previous studies have demonstrated that LLM-based judges can effectively match both controlled and crowdsourced human preferences in various evaluation tasks \cite{zheng2023judgingllmasajudgemtbenchchatbot}. With this approach, when evaluating any model outputs for questions in Tele-Eval, we prompt Mixtral 8x-7B-Instruct to compare the model's output to the ground truth and provide a Yes or No answer regarding its correctness. We define the LLM-Eval metric as the average score of these boolean responses, such that
\begin{equation}
\text{LLM-Eval} = 100\times \frac{1}{N} \sum_{i=1}^{N} \mathbbm{1}\left\{\text{LLM judges $\boldsymbol{y}^i$ as correct}\right\},
\end{equation}
where $\boldsymbol{y}^i$ is the model's answer to question $i$ of Tele-Eval, $N$ is the total number of evaluated outputs, and $\mathbbm{1}\{\cdot\}$ is the indicator function.

\section{Initial Experiments}
With both the training and evaluation datasets prepared, the next step is to train our models. However, before proceeding, several key questions regarding training techniques and parameters need to be addressed to lay the foundation for the training procedure. While general guidelines on training parameters, such as learning rate and batch size, can be drawn from previous research \cite{labrak2024biomistralcollectionopensourcepretrained, colombo2024saullm7bpioneeringlargelanguage, xie2023efficientcontinualpretrainingbuilding}, the characteristics of telecommunications knowledge—such as the prevalence of equations—introduce specific challenges. Key questions that need to be answered include whether full parameter fine-tuning is necessary or if parameter-efficient tuning methods are sufficient, as well as determining the number of training epochs needed to effectively adapt the LLM to the telecommunications domain. We address these questions in this section, while noting that our results can serve as a guide for adapting other models to the telecommunications domain beyond those discussed in our paper.
\subsection{Training Settings}
\label{sec:trainingsettings}
For these experiments, we considered two models: Gemma-2B and LLaMA-3-8B. This choice was motivated by our intention to explore the behavior of model training across different parameter sizes. Throughout these studies, we use 20\% of the Tele-Data dataset as the training data, with 10\% of this subset serving as the validation set. To solve eq. (\ref{eq:continual}), we set the batch size to 4M tokens (with a sequence length of 8192 tokens) and use the AdamW optimizer with a weight decay of 0.1, keeping the default Hugging Face $\beta$ parameters unchanged. The maximum gradient norm is set to $1$. The learning rate is decreased according to a cosine learning rate schedule, down to 10\% of the maximum learning rate, and a linear warmup of 10\% of an epoch is applied \cite{gupta2023continual}.
The maximum learning rate is set to 1e-5 for LLaMA and 2e-5 for Gemma. These values are aligned with the pretraining stages of these LLMs \cite{dubey2024llama3herdmodels, gemmateam2024gemmaopenmodelsbased}. To improve efficiency, we employ mixed-precision training alongside sample packing and block attention techniques to avoid cross-contamination \cite{krell2022efficientsequencepackingcrosscontamination}.
\subsection{PEFT vs. FFT}
\begin{figure*}[t]
    \centering
    \includegraphics[width=0.99\textwidth]{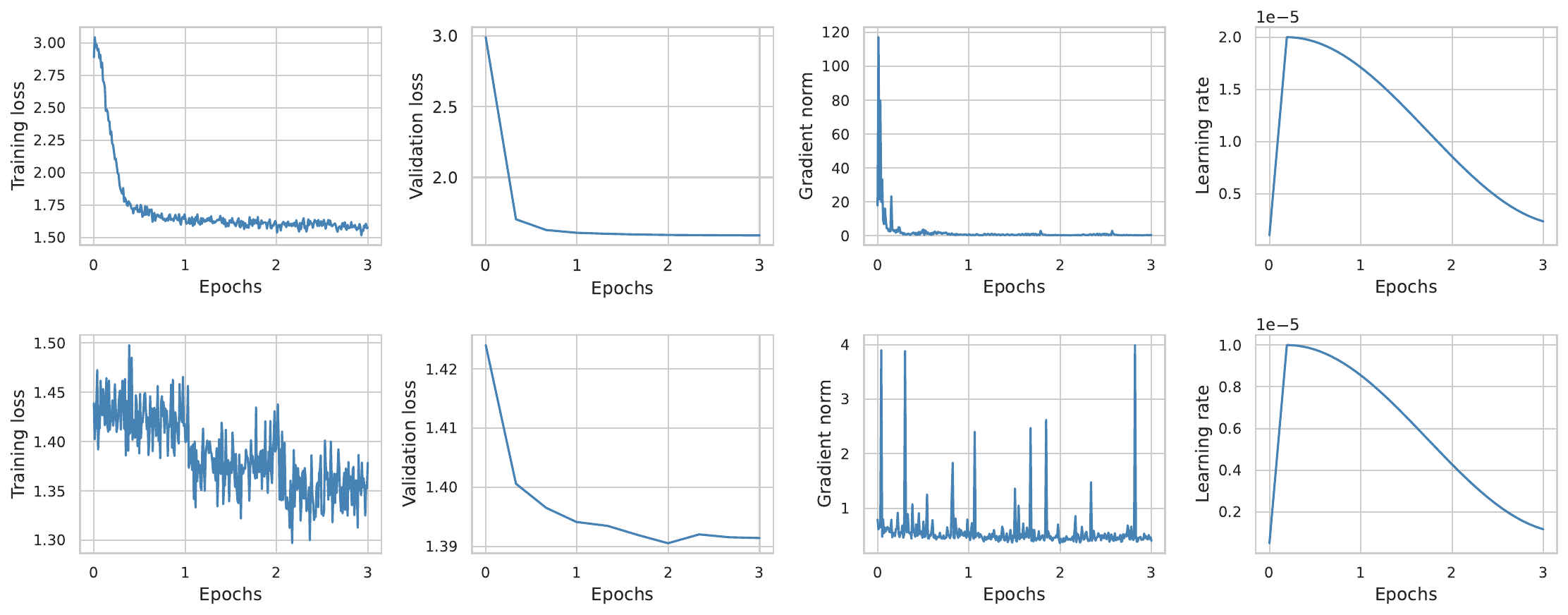}
    \caption{Training metrics for Gemma-2B (top) and Llama-3-8B (bottom) models across three epochs.}
    \label{fig:threeepochs}
\end{figure*}
PEFT techniques aim to reduce computational costs and memory requirements during domain adaptation by training only a minimal number of parameters, in contrast to FFT methods that update all the model's parameters. Among these techniques, LoRa \cite{hu2022lora} stands out as a prominent PEFT method. It adds small, trainable low-rank matrices to the existing weights of the pretrained model. To investigate whether PEFT methods like LoRa are sufficient for adapting language models to the telecommunications domain, we conducted a comparative study between LoRa and FFT using both Gemma-2B and LLaMA-3-8B models. For our LoRa implementations, we fixed the LoRa rank to r=64, set the Lora alpha to 32, and used a LoRa dropout rate of 0.1.

The results, reported in Fig. \ref{fig:peftvsfft}, demonstrate that for smaller models, LoRa can initially inject telecom knowledge but quickly saturates due to its limited capacity. In contrast, for the LLaMA-3-8B model, which is more knowledgeable about telecommunications, the gradient norm of the LoRa method remained extremely low. This hindered parameter updates, causing the training loss to barely change and limiting LoRa's ability to inject additional knowledge. These findings indicate that, despite its computational efficiency, LoRa may not be sufficient to adapt the model to the telecommunications field. Consequently, in the next sections,  we will rely on full parameter fine-tuning for all our training.
\subsection{One or Multiple Epochs?}
\label{subsection:oneormultiple}
A crucial question when adapting an LLM to a target domain, in our case being telecommunications, is determining the most effective duration for model training. Is a single epoch enough to adapt the model, or do multiple exposures to the data with shuffling at each epoch yield better results? To address this, we conducted experiments using two different model scales: Gemma-2B and LLama-3-8B. As we will demonstrate, this distinction arises from the distinct behaviors of these models due to their different sizes. The results are reported in Fig. \ref{fig:threeepochs} and discussed below.

\xhdr{Gemma-2B} As shown in Fig. \ref{fig:threeepochs}, the training loss of Gemma-2B demonstrates consistent behavior, decreasing and then plateauing after approximately two epochs. A similar pattern is observed in the validation loss, which also stabilizes after two epochs. Additionally, the gradient norm approaches 0, indicating that the training is reaching convergence. It is worth noting that for smaller models, after three epochs of training, there are no signs of overfitting, as both validation and training losses plateau at similar levels. This suggests that the model has developed a comprehensive understanding of the telecom data, and further training results in only minor adjustments to this understanding without significantly adding to or diminishing the existing telecom knowledge.

\xhdr{LLama-3-8B} The first observation one can make by investigating the bottom plots of Fig. \ref{fig:threeepochs} is that the training loss starts with a cross-entropy of around 1.42. This indicates that the model is already quite knowledgeable when it comes to telecommunications. This is not surprising for two reasons: first, this model was trained on 15T tokens, hence it has likely encountered a wide variety of telecom-related data; second, the large number of parameters in the model allows it to retain such information. Regardless, continual pretraining is helpful to further solidify this knowledge. 

By further examining the training loss, an interesting phenomenon can be witnessed: at the end of each epoch, the training loss experiences a noticeable drop. This suggests that the model is memorizing patterns in the training dataset and becoming more confident in its next token prediction as it encounters these patterns after each epoch. These patterns have some generalizable components, as evidenced by the drop in validation loss up to the second epoch. However, beyond two epochs, this memorization becomes problematic as the learned patterns become more specific to the training data and less transferable to the validation data, thus leading to overfitting. We hypothesize that two epochs represent a sweet spot: the first epoch serves to warm up the model and absorb telecom knowledge, while the second epoch, aided by the cosine decay, shapes this knowledge with a very low learning rate, allowing the model to make final adjustments in the loss landscape. Based on the above, we use two epochs to create the Tele-LLMs series in the next section.

\begin{remark} We have observed similar training loss behavior across other models (e.g., Mistral-7B). As the number of parameters increases, the model becomes more expressive and better able to capture intricate relationships in the training data, leading to this behavior. We have not reported these results due to their similarity to those shown in the bottom plots of Fig. \ref{fig:threeepochs}.
\end{remark}

\section{Tele-LLMs}
\label{sec:trainingsection}
\begin{figure*}[t]
    \centering
    \includegraphics[width=0.99\textwidth]{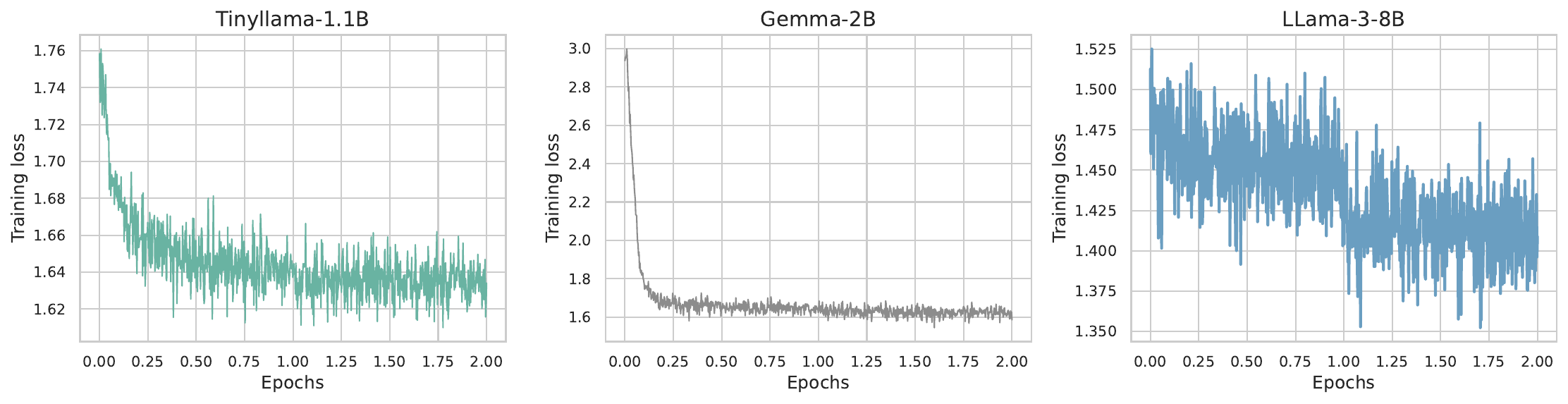}
    \caption{Training loss of the Tele-LLMs series.}
    \label{fig:finalepoch}
\end{figure*}
\begin{table*}[t]
\centering
\begin{tabular}{l|ccc|ccc}\toprule
 & \multicolumn{3}{c|}{\textbf{Tele-Eval}} & \multicolumn{3}{c}{\textbf{General Knowledge}} \\

 & \multicolumn{1}{c}{Ans-PPL} & \multicolumn{1}{c}{SemScore} & \multicolumn{1}{c|}{LLM-Eval} &\multicolumn{1}{c}{MMLU} & \multicolumn{1}{c}{HellaSWAG} & \multicolumn{1}{c}{GSM8K} \\ \midrule
Tinyllama-1.1B & 230.44 & 0.5952 & 8.26 & 0.2501 & \textbf{0.4670} & \textbf{0.0258} \\
Tinyllama-1.1B-Tele  & \textbf{9.06} & \textbf{0.6093} & \textbf{11.37} & \textbf{0.2519} & 0.4663 & 0.0212 \\
\midrule
Gemma-2B & 13.31 & 0.5998 & 13.59 & 0.3289 & 0.5275 & 0.0576 \\
Gemma-2B-Tele & \textbf{11.73} & \textbf{0.6302} & \textbf{17.07} & \textbf{0.3497} & \textbf{0.5304} & \textbf{0.0622} \\
\midrule
Gemma-2-2B & 12.96 & 0.5853 & 15.10 & \textbf{0.4789} & 0.5164 & 0.0607 \\
Gemma-2-2B-Tele & \textbf{10.86} & \textbf{0.6237} & \textbf{17.34} & 0.4773 & \textbf{0.5212} & \textbf{0.0697} \\
\midrule
LLama-3.2-1B  & 11.72 & 0.6205 & 11.85 & 0.3611 & \textbf{0.4769} & 0.0288 \\
LLama-3.2-1B-Tele & \textbf{10.50} & \textbf{0.6318} & \textbf{12.98} & \textbf{0.3631} & 0.4751 & \textbf{0.0303}\\
\midrule
LLama-3.2-3B   & 9.52 & 0.6413 & 24.68 & \textbf{0.5409} & \textbf{0.5522} & \textbf{0.1152} \\
LLama-3.2-3B-Tele & \textbf{9.16} & \textbf{0.6557} & \textbf{26.71} & 0.5366 & 0.5489 & 0.1008\\
\midrule
LLama-3-8B  & 9.17 & 0.6358 & 24.60 & \textbf{0.6209} & 0.6009 & 0.144 \\
LLama-3-8B-Tele & \textbf{8.49} & \textbf{0.6482} & \textbf{29.60} & 0.6157 & \textbf{0.6068} & \textbf{0.2441}\\
\bottomrule
\end{tabular}
\caption{LLMs' performance across Tele-Eval and other general knowledge benchmarks.}
\label{tab:trainingresults}
\end{table*}


Having completed the initial experiments, we proceed with the full-scale training of the LLMs to create the telecom-adapted series of models, starting with the following models:
TinyLlama-1.1B, Gemma-2B, Gemma-2-2B, LLaMA-3.2-1B, LLaMA-3.2-3B, and LLaMA-3-8B. Our selection of these base models was guided by a combination of factors, including their performance on the Hugging Face Open LLM leaderboard and a consideration of their licenses. We also accounted for the diversity of the releasing companies to mitigate potential risks associated with changes in these licenses.

\xhdr{Training Settings} We maintain the training settings described in Section \ref{sec:trainingsettings}, where models smaller than 8B parameters follow the same approach as Gemma-2B while aligning the sequence length to each model's pretraining settings. For the training dataset across the entire series, we utilize the entire Tele-Data dataset for two epochs, augmented with 5\% of regular general-purpose data taken from SlimPajama\footnote{https://huggingface.co/datasets/cerebras/SlimPajama-627B}. This augmentation is designed to mitigate the catastrophic forgetting phenomenon by reintroducing general-purpose data to the model, helping it retain key skills acquired during the initial pretraining stage. The training of these models was conducted on a cluster of 8 NVIDIA A6000 GPUs, consuming a total of 5,000 GPU hours.

\xhdr{Loss Trends} In Fig. \ref{fig:finalepoch}, we report the training loss across three models. Note that all sub-8B parameter models 
exhibited similar trends to those of Gemma-2B and Tinyllama-1.1B shown in Fig. \ref{fig:finalepoch}. These models, in particular, demonstrate steady training loss behavior, while LLaMA-3-8B displays the sharp loss reduction characteristic of larger models at each epoch depicted in the previous section.


\xhdr{Evaluation Settings} We evaluate the LLMs on Tele-Eval using the three metrics discussed in Section \ref{sec:evaluationmetrics}, generating model answers through greedy decoding of 100 new tokens with the prompt provided in Appendix \ref{appendix:prompts}. To assess the impact of our training on other LLM capabilities, we evaluate our models on three additional benchmarks: MMLU \cite{hendrycks2021measuring} for general world knowledge, HellaSWAG \cite{zellers-etal-2019-hellaswag} for commonsense understanding, and GSM-8K \cite{cobbe2021trainingverifierssolvemath} for mathematical reasoning abilities. All evaluations, including Tele-Eval and the additional benchmarks, are conducted in a zero-shot setting.

\xhdr{Quantitative Evaluation} 
As shown in Table \ref{tab:trainingresults}, the adapted telecom LLMs outperform their counterparts across all metrics on Tele-Eval. For general knowledge tasks, these models exhibit stable performance compared to the original models, with slight variations—either improvements or degradations—across most tasks. Notably, there are improvements on GSM-8K, with LLama-3-8B-Tele achieving a 10\% higher accuracy. This enhancement is attributed to the equations found in Tele-Data, which enable the adapted model to become stronger in mathematical reasoning. 

Comparing the LLMs' performance on Tele-Eval across the three metrics reveals several insights: First, Ans-PPL shows minimal relative differences for larger models. This occurs because larger models, with their higher proficiency, are less surprised by the terms used in the ground truth answers. Consequently, improvements in telecommunications knowledge become less apparent and more diluted in this metric. Another observation is the lack of comparability across different models using the Ans-PPL metric, as evidenced by the performance comparison between Tinyllama-1.1B-Tele and Gemma-2B-Tele. The tendency of a model to follow certain formats when answering questions biases this metric, resulting in either better or worse performance. This makes Ans-PPL more useful for comparing models within the same family rather than across different families, due to the heterogeneous nature of their tendencies.

SemScore exhibits patterns similar to Ans-PPL; an LLM producing coherent responses with appropriate technical terminology may achieve good scores even when providing incorrect answers. With this in mind, and given the heterogeneous response styles across LLM families, both SemScore and Ans-PPL are more effective at highlighting significant differences within the same LLM family. For detecting subtle disparities and comparing across different LLM families, LLM-Eval proves more robust. It disregards linguistic variations and instead focuses on the underlying concepts being assessed. Using this metric, we observe an average relative improvement of 25\% for the telecom-adapted LLMs.

Although LLM-Eval provides the most accurate evaluation, its drawback is the runtime. Given the $O(n^2)$ scaling of the attention mechanism, where $n$ is the input prompt length for the evaluation, our evaluation with Mixtral-8x-7B took approximately 50 milliseconds per question on a single Nvidia A6000 GPU. Note that given the size of Tele-Eval, the standard deviation of our results is negligible. 


\begin{remark}
    It is worth noting that Tele-Eval, with its focus on highly granular telecommunications concepts, presents a significant challenge for LLMs, as evidenced by their performance on the LLM-Eval metric. This inherent difficulty is one of Tele-Eval's advantages; if the questions were easily answerable, the dataset would fail to effectively distinguish the advantages gained through adapting models to the telecom domain.
\end{remark}

\xhdr{Qualitative Evaluations} Going beyond the quantitative comparison provided in Table \ref{tab:trainingresults}, we offer a set of examples in Appendix \ref{sec:appendixqualitative} that illustrate how the behavior of these models change, using Gemma-2B and Gemma-2B-Tele as a case study. These examples highlight the model's enhanced ability to provide detailed explanations, confidently answer telecommunications-related questions, and even delve into telecommunications concepts when they are mentioned in the prompt. These qualitative observations further emphasize the successful adaptation of these models to the telecommunications domain.

\section{Further Exploration}
In the final phase of our work, we investigate various adaptation strategies and specific dynamics that emerge when tailoring LLMs to the telecommunications domain. We particularly focus on two key aspects: the potential for expertise division in the adaptation process, and the unique adaptation dynamics that arise based on each model's characteristics and training data. We also adapt our models to follow instructions, creating chatbot-like models that users can interact with. We show that these adapted models demonstrate greater proficiency in telecommunications compared to their general-purpose instruct counterparts.
\subsection{Division of Expertise}
\label{sec:divisionexpertise}
\begin{table}[t]
\centering
\resizebox{\linewidth}{!}{
\begin{tabular}{l|cc|ccc}\toprule
 & \multicolumn{2}{c|}{\textbf{Tele-Eval}} & \multicolumn{3}{c}{\textbf{General Knowledge}} \\
& \multicolumn{1}{c}{Standards} & \multicolumn{1}{c|}{Overall} &\multicolumn{1}{c}{MMLU} & \multicolumn{1}{c}{HellaSWAG} & \multicolumn{1}{c}{GSM8K} \\ \midrule
Gemma-2B & 8.12 &  13.59 & 0.3289 & 0.5275 & 0.0576 \\
Gemma-2B-Tele  &  \textbf{10.15} & \textbf{17.07} & \textbf{0.3497} & \textbf{0.5304} & \textbf{0.0622} \\
Gemma-2B-Standards  &  9.58 & 11.18 & 0.3314 & 0.5250 & 0.0432 \\
\bottomrule
\end{tabular}}
\caption{Performance comparison between Gemma-2B, Gemma-2B-Tele, and Gemma-2B-Standards.}
\label{tab:divisionexpertise}
\end{table}
When adapting an LLM to the telecommunications domain, a key question arises: should we adapt a single LLM to cover the entire domain, or should we adapt multiple LLMs, each specialized in a specific aspect of telecommunications? For example, one LLM could focus on scholarly material while another concentrates on standards. To explore this, we compare two versions of the Gemma-2B model: one trained exclusively on the standards portion of Tele-Data, referred to as Gemma-2B-Standards, and Gemma-2B-Tele. The training parameters align with those previously reported for Gemma-2B-Tele. Evaluation is based on both the entirety of Tele-Eval and questions related to standards using the LLM-eval metric, along with general knowledge evaluation datasets to assess the retention of the model's capabilities. The results are reported in Table \ref{tab:divisionexpertise}.

As shown, Gemma-2B-Tele outperforms Gemma-2B-Standards on the standards-related questions while also demonstrating stronger general knowledge capabilities. Additionally, Gemma-2B-Standards performs worse than the base model Gemma-2B on the overall Tele-Eval dataset. The reason behind this is that the Gemma-2B-Standards model, having been trained solely on standards, becomes more attuned to that specific type of content. Standards tend to be highly technical with a unique token distribution that differs from the broader telecommunications knowledge found in scholarly and Wikipedia articles. This narrow focus makes the model less capable of handling scholarly material compared to the base model, leading to underperformance on Tele-Eval. At the same time, this narrow focus also negatively impacts the LLM's general abilities. All in all, this highlights that the most effective strategy is to adapt the LLM to the entire telecommunications dataset, thereby benefiting from the transfer learning that occurs across these diverse materials rather than narrowly focusing on a single aspect.

\subsection{Pretraining Data Impact}
\begin{figure}[t]
\centering
\begin{tcolorbox}[colframe=niceblue!75!black, colback=blue!5!white, sharp corners=south, title=Instructions Following]
\textbf{Prompt:} A poem about telecommunications.

\medskip

\textbf{Phi-1.5:} Answer: Telecommunications is a field that connects people and places,
Through wires and signals, it's a vital task.
It's like a bridge that spans the distance,
Bringing us closer, no matter the distance.

\medskip

\textbf{Gemma-2B:} The poem is about the telephone. It is about the telephone that is in the room. 
\end{tcolorbox}
\caption{Generated responses from different models for a telecommunications-themed prompt.}
\label{fig:prompt-trial}
\end{figure}

Perhaps the most impactful element that shapes an LLM's behavior is its pretraining data. This data is generally not publicly available, so one can only infer its type through interactions with the model. Understanding this behavior is crucial because when adapting an LLM to the telecom domain, specific trends emerge based on this behavior.

To illustrate this, let us consider two models: Gemma-2B and Gemma-2B-it\footnote{\url{https://huggingface.co/google/gemma-2b-it}}. The latter is an instruct version of the former that has been post-trained on instruction-following data, which includes tasks like question answering and summarization. As Table \ref{tab:pretrainingimpact} shows, instruct models outperform their base counterparts on Tele-Eval. This is because the additional instruction-focused training allows the LLM to better handle questions, follow instructions, and leverage its internalized knowledge to effectively answer queries. This advantage is significant enough that Gemma-2B-it is able to outperform our telecommunications-adapted LLM, Gemma-2B-Tele. However, since Gemma-2B-Tele has been enriched with telecommunications knowledge, we can surpass the performance of Gemma-2B-it by applying a similar post-training instruction-following adaptation using a dataset like Alpaca\footnote{\url{https://huggingface.co/datasets/tatsu-lab/alpaca}} and utilizing the same training settings previously detailed, thus creating Gemma-2B-Tele-Alpaca.

\begin{table}[t]
\centering
\begin{tabular}{l|ccc}\toprule
& \multicolumn{3}{c}{\textbf{Tele-Eval}} \\
& \multicolumn{1}{c}{Original} & \multicolumn{1}{c}{Tele} &\multicolumn{1}{c}{Tele-Alpaca} \\ \midrule
Gemma-2B  & 13.59 &  17.07 & 25.31  \\
Gemma-2B-it  & 19.84 &  18.05 & 24.83  \\
Phi-1.5  &  14.87 & 13.06 & 18.84  \\
\bottomrule
\end{tabular}
\caption{LLM-Eval performance on Tele-Eval.}
\label{tab:pretrainingimpact}
\end{table}

The significance of the above observation is that some base models have been trained on instruct-like data during their pretraining stage. An example of this is Microsoft Phi-1.5. In their paper \cite{gunasekar2023textbooksneed}, the authors specify that Phi-1.5 can be prompted as an instruct model because the pretraining data included a large portion of question-answer formats. This leads to noticeable behavioral differences between Gemma-2B and Phi-1.5, as illustrated in Fig. \ref{fig:prompt-trial}. One can see that Phi-1.5 behaves more like a chatbot, following the instructions of the prompt, while Gemma-2B fills in sentences without treating the input prompt as an instruction to follow.

Given the above, when applying the training recipe described in our paper to Phi-1.5, the performance on Tele-Eval declines for Phi-1.5-Tele, similar to what occurred for Gemma-2B-it. However, by post-training Phi-1.5-Tele on Alpaca using the same training settings, the performance significantly improves beyond that of the base model. This is because, by doing so, we align the model back to its original behavior. With this in mind, it is crucial to be mindful of the model's behavior before proceeding with the adaptation process, in order to anticipate its dynamics accordingly.

Given the above observations and the special behavior of the Phi-1.5 model, we only release the Alpaca fine-tuned version of Phi-1.5 as part of our Tele-LLM series. It is worth noting that its performance on general knowledge datasets is 0.3965 on MMLU, 0.4680 on HellaSWAG, and 0.1107 on GSM8K, compared to 0.4072, 0.4798, and 0.0576, respectively, for the original Phi-1.5. This demonstrates that our model performs better in telecom while still retaining its general knowledge and even gaining in mathematical performance.

\begin{table}[t]
\centering
\resizebox{\linewidth}{!}{
\begin{tabular}{l|cccc}\toprule
& \multicolumn{4}{c}{\textbf{Tele-Eval}} \\
& \multicolumn{1}{c}{Base} & \multicolumn{1}{c}{Instruct}  & \multicolumn{1}{c}{Tele} &\multicolumn{1}{c}{Tele-Instruct} \\ \midrule
Tinyllama-1.1B  & 8.26 & 15.42 &  11.37 & 17.40  \\
Gemma-2B  & 13.59 & 19.84 &  17.07 & 27.78  \\
Gemma-2-2B  & 15.10 & 22.30 &  17.34 & 25.46  \\
LLama-3.2-1B  & 11.85 & 14.50 &  12.98& 16.20 \\
LLama-3.2-3B  & 24.68 & 25.20 & 26.71& 28.56 \\
LLama-3-8B  & 24.60 & 30.65 &  29.60 & 34.51 \\
\bottomrule
\end{tabular}}
\caption{LLM-Eval performance on Tele-Eval.}
\label{tab:instructperf}
\end{table}
\subsection{Instructions Fine-tuning}
\begin{table*}[]
\caption{Performance comparison of individually tuned models vs. unified instruction tuning}
\centering
\resizebox{\linewidth}{!}{
\begin{NiceTabular}{c|c|c|cc|cc|cc|cc} 
\toprule 
\textbf{Models}
&\textbf{Citation Link Prediction}
&\textbf{Citation Recommendation}
&\multicolumn{2}{c|}{\textbf{Title Generation}}
&\multicolumn{2}{c|}{\textbf{Abstract Completion}}
&\multicolumn{2}{c|}{\makecell{\textbf{Citation Sentence} \\ \textbf{Generation}}}
&\multicolumn{2}{c}{\makecell{\textbf{Introduction} \\ \textbf{to Abstract}}} \\
\midrule
& Accuracy & Accuracy & Precision & F1 Score & Precision & F1 Score & Precision & F1 Score & Precision & F1 Score \\
\midrule 
 Llama-3.2-1B &21.34 &10.04&0.7886  &0.8035  &0.8353 &0.8184 &0.8065 &0.7941  &0.7994 &0.7944  \\
 Llama-3.2-1B-Lit  &51.26 &56.1 &0.7930 &0.8279  &0.8494 &0.8296  &0.8267 &0.8216 &0.8300  &0.8314  \\
   Llama-3.2-1B-Tele-Lit &52.61 &58.29 &0.7978 &0.8294  &0.8519 &0.8508 &0.8316  &0.8286  &0.8318 &0.8326  \\

\midrule
 Llama-3.2-3B &39.29 &25.5 &0.8050 &0.8207 &0.8337 &0.8282  &0.7991 &0.7904  &0.8073 &0.8040  \\
 Llama-3.2-3B-Lit  &56.56 &61.45 &0.8049 &0.8381 &0.8486  &0.8512  &0.8331 &0.8289  &0.8291 &0.8315  \\
  Llama-3.2-3B-Tele-Lit&57.76 &64.61 &0.8097 &0.8409  &0.8479 &0.8345 &0.8309  &0.8252  &0.8314 &0.8336  \\

\midrule
 Llama-3-8B &51.75 &55.67 &0.8091 &0.8171  &0.8456 &0.8240 &0.8123  &0.7998  &0.8076 &0.8066  \\
  Llama-3-8B-Lit &75.56 &63.6 &0.7903 &0.8262  &0.8548 &0.8570 &0.8343  &0.8306  &0.8332 &0.8338  \\
 Llama-3-8B-Tele-Lit &78.56 &66.98 &0.7942 &0.8286 &0.8580 &0.8605  &0.8338 &0.8317  &0.8314 &0.8330  \\
\midrule
 GPT-4o &79.85 &64.55 &0.8648 &0.8764  &0.8544 &0.8483 &0.8259  &0.8116  &0.8723 &0.8457 \\
\bottomrule
\end{NiceTabular}}
\label{tab:ablation_tasks}
\end{table*}
Although base models are essential, as they contain the raw LLM's knowledge and are best suited for fine-tuning for specific telecommunications applications, it is common for users to interact with these models as chatbots. Therefore, in this section, we proceed to fine-tune our telecom-adapted models to follow instructions through instructions fine-tuning.

\xhdr{Training Settings} We maintain the training settings from Section \ref{sec:trainingsection} with minor adjustments. Specifically, we decrease the batch size to 128k tokens, set the context length to 2048, and limit the number of epochs to 1. For this stage, we use two datasets: Alpaca and Open-Instruct\footnote{\url{https://huggingface.co/datasets/VMware/open-instruct}}. The combined datasets provide approximately 200k samples for training.

\xhdr{Results} 
In Table \ref{tab:instructperf}, we compare the Tele-Instruct versions of our models to their general instruct counterparts. Our instruction-tuned series outperforms the general instruct models, thereby showcasing their superior knowledge in telecom compared to their general-purpose counterparts.

\subsection{Benefits of Specialization}
A central question we explore is how well domain specialization improves performance on downstream tasks beyond question answering. To investigate this, we focus on several literature-centric tasks: citation prediction, citation recommendation, title generation, abstract completion, and introduction-to-abstract conversion. Leveraging the Tele-Data dataset, we curate fine-tuning samples tailored to each of these tasks. We then fine-tune base models from the Llama family and compare their performance to the specialized Tele-LLMs, using a dedicated hold-out set excluded from training. Details on the fine-tuning process—including instruction datasets and hyperparameters—are provided in Appendix \ref{appendix:datasetprep}.

For tasks involving text generation (e.g., citation sentence generation or abstract-to-title conversion), we adopted BERTScore-incorporating both precision and F1 measures- as our primary evaluation metric \cite{zhang2020bertscoreevaluatingtextgeneration}. As shown in Table \ref{tab:ablation_tasks}, fine-tuned models (referred to by the suffix Lit) has the potential of outperforming GPT-4o on literature tasks (e.g., as is the case for LLama-3-8B) where domain-specific knowledge and graph structure are critical, while GPT-4o maintained an advantage in English language formulation. Notably, these fine-tuned models achieve competitive performance despite being significantly smaller than GPT-4o, demonstrating the efficiency benefits of targeted fine-tuning. Furthermore, the fine-tuned telecom models (referred to by the suffix Tele-Lit) consistently outperformed their generic counterparts across all tasks. This demonstrates how domain specialization fundamentally improves the model's representations, aligning them more effectively with the telecom domain and enhancing performance on downstream tasks after task-specific fine-tuning. These results highlight the value of our developed Tele-LLMs series in enhancing performance on telecom-specific downstream tasks.

\section{Conclusions}
In this paper, we addressed the challenge of adapting LLMs for specialized use in telecommunications. In our endeavor, we created and released Tele-Data and Tele-Eval, a comprehensive telecommunications training and evaluation datasets. Through extensive experimentation, we identified the most suitable training strategies for adapting LLMs to this domain. The culmination of our work is Tele-LLMs, a series of open-source models ranging from 1B to 8B parameters, specifically designed for the telecommunications domain. These models outperform their general-purpose counterparts on Tele-Eval while retaining their broader capabilities. Beyond this culmination, our work also investigated various adaptation strategies, such as the division of expertise, and explored the dynamics that arise during the adaptation process depending on the models involved. As a future direction, our work will aim to leverage Tele-LLMs and augment them with multi-modal capabilities to understand and reason about wireless measurements and signals.
\bibliographystyle{IEEEtran}
\bibliography{references}

\section*{Appendices}
\appendices
\section{Manuscripts Cleaning Procedure}
\label{appendix:cleaning}
Given the importance of ensuring that the arXiv papers are clean for training, we employ a rigorous cleaning process on these sources. Our process begins with the removal of all comments written in the LaTeX files. For this, we leverage Google's arXiv LaTeX Cleaner\footnote{\url{https://github.com/google-research/arxiv-latex-cleaner}}.

Next, because LaTeX sources can include multiple LaTeX files, we start by unifying the LaTeX commands used for importing these files. Particularly, we ensure that imports use the \textbackslash input\{$\cdot$\} command. By creating a directed graph that maps these relationships, we identify the main LaTeX file of the paper. We then use the Latexpand Perl script\footnote{We tested fatex (\url{https://ctan.org/pkg/fatex}) and fap (\url{https://github.com/fchauvel/fap}), but achieved the best results with latexpand ( \url{https://ctan.org/pkg/latexpand}).} to flatten the document, ensuring the main file contains all material.

In the subsequent step, we address the author's custom commands by `de-macoring' them. To do so, we unify all custom commands—from \textbackslash def\{$\cdot$\} and \textbackslash DeclareMathOperator\{$\cdot$\} to \textbackslash newcommand\{$\cdot$\}—before using the Python library de-macro (\url{https://ctan.org/pkg/de-macro}) to replace these macros with their native LaTeX equivalents. Afterwards, we target the removal of figures and tables, as our focus is on the in-line text and equations. We also compile a list of over a hundred LaTeX commands and environments that are not informative, which we remove using regular expression matching. This ensures that the final LaTeX files contain only the text and equations without any LaTeX residue.

Additionally, given the many forms that citations can take, along with the wide variety of text within their brackets, we unify all citation, label, and reference commands. This standardization helps the LLM to avoid dealing with heterogeneity during training. Lastly, we remove all preambles of the LaTeX files and ensure that the title of the manuscript is retained. This provides a unified format for these sources, thus facilitating the LLM training in the next stage.

\section{Samples of Tele-Data}
\label{appendix:teledata}
We provide below an example of each category of Tele-Data. The [...] symbol is inserted below to reduce the size of the strings.\\ \\
\textbf{ID:} arxiv\_14326 \\
\textbf{Category:} \color{niceblue} \textbf{arxiv} \color{black} \\
\textbf{Content:} Flexible-Position MIMO for Wireless Communications: Fundamentals, Challenges, and Future Directions\textbackslash n \textbackslash n Abstract\textbackslash n \textbackslash n The flexible-position multiple-input multiple-output (FLP-MIMO), such as fluid antennas and movable antennas, is a promising technology for future wireless communications [...]\\
\textbf{Metadata:}\\
\indent \textbf{Arxiv\_id:} 2308.14578\\ 
\indent \textbf{Title:} Flexible-Position MIMO for Wireless Communications: Fundamentals, Challenges, and Future Directions\\
\indent \textbf{Abstract:} The flexible-position multiple-input multiple-output (FLP-MIMO), such as [...]
\\ 

\noindent\textbf{ID:} standard\_2413 \\
\textbf{Category:} \color{niceblue} \textbf{standard} \color{black} \\
\textbf{Content:} 3rd Generation Partnership Project; \textbackslash n Technical Specification Group Core Network and Terminals;\textbackslash n Interworking between the Public Land Mobile Network (PLMN)\textbackslash n supporting packet based services with\textbackslash n Wireless Local Area Network (WLAN) Access and\textbackslash n Packet Data Networks (PDN)\textbackslash n (Release 12)\textbackslash n Foreword\textbackslash n This Technical Specification (TS) has been produced [...] \\
\textbf{Metadata:}\\
\indent \textbf{Series:} 29\\ 
\indent \textbf{Release:} 12\\
\indent \textbf{File\_name:} 29161-c00
\\ 

\noindent\textbf{ID:} wiki\_5438 \\
\textbf{Category:} \color{niceblue} \textbf{wiki} \color{black} \\
\textbf{Content:}A backbone or core network is a part of a computer network which interconnects networks, providing a path for the exchange of information between different LANs or subnetworks. A backbone can tie together diverse networks [...] \\
\textbf{Metadata:}\\
\indent \textbf{Title:} Backbone network\\ 
\indent \textbf{Url:} https://en.wikipedia.org/wiki/Backbone\%20network\\

\noindent\textbf{ID:} web\_71187 \\
\textbf{Category:} \color{niceblue} \textbf{web} \color{black} \\
\textbf{Content:}1. Field of the Invention\textbackslash n The present invention relates generally to methods of addressing data packets destined to a host in a communications network, and particularly to a method of defining an address for a mobile terminal/host [...] \\
\textbf{Metadata:}\\
 \indent \textbf{Url:} http://www.google.com/patents/US6147986

\section{LLM Prompts}
\label{appendix:prompts}
The following prompts were utilized throughout our framework to develop and evaluate various areas. In summary, and in order, the prompts were used for the following tasks:
\begin{enumerate}
\item \textbf{Prompt 1:} Used to find arXiv papers related to the telecommunications and networking domains.
\item \textbf{Prompt 2:} Used to find websites and Wikipedia pages related to the telecommunications and networking domains.
\item \textbf{Prompt 3:} Leveraged to generate the QnAs that form the initial Tele-eval dataset.
\item \textbf{Prompt 4:} Leveraged to filter out locally relevant QnAs, resulting in the filtered Tele-eval dataset.
\item \textbf{Prompt 5:} Used to instruct the base model to complete the answer based on the provided question.
\item \textbf{Prompt 6:} Utilized as a prompt for LLM-Eval.  
\end{enumerate}
\bigskip
\begin{lstlisting}[style=promptstyle, caption={arXiv filtering}, basicstyle=\ttfamily]
Given the following scientific paper abstract: {Abstract}, Answer by Yes or No if this paper is related to the telecommunications and networking domains.
\end{lstlisting}
\bigskip
\begin{lstlisting}[style=promptstyle, caption={Wiki \& websites filtering}, basicstyle=\ttfamily]
Given the following website content: {Website}, Answer by Yes or No if this content contains technical content about the telecommunications and networking domains.
\end{lstlisting}
\clearpage
\begin{lstlisting}[style=promptstyle, caption={Evaluation dataset generation}, basicstyle=\ttfamily]
Generate 5 questions and short answers based on the following passage: {Passage}. 
The questions should follow this format:
Question 1: What frequency band does Bluetooth use?
Answer 1: 2.4 GHz
Question 2: Which pairs of wires are used in 10/100Base-T?
Answer 2: Pair 2 and pair 3
Question 3: What is a Heterogeneous Network?
Answer 3: A network consisting of multiple cells with
different characteristics
\end{lstlisting}
\bigskip
\medskip
\begin{lstlisting}[style=promptstyle, caption={Questions filtering}, basicstyle=\ttfamily]
Answer by Yes or No if I can find the answer to this question online without access to the material from which the question came: {question}
\end{lstlisting}
\bigskip
\medskip
\begin{lstlisting}[style=promptstyle, caption={Answering format}, basicstyle=\ttfamily]
The following is a question about telecommunications and networking.
Question: {{statement}}
Answer:
\end{lstlisting}
\bigskip
\medskip
\begin{lstlisting}[style=promptstyle, caption={LLM-Eval format}, basicstyle=\ttfamily]
Task: Evaluate the correctness of a provided answer to a telecommunications and networking question. 
Question: {question}
Ground Truth Answer: {ground_truth}
Provided Answer: {prediction}
Instructions:
1. Compare the Provided Answer to the Ground Truth Answer.
2. Determine if the Provided Answer is correct based on the Ground Truth Answer.
3. Respond with only Yes or No.
Is the Provided Answer correct?
\end{lstlisting}
\bigskip
\medskip

\section{Qualitative Examples}
\label{sec:appendixqualitative}
Below, we present examples that illustrate the qualitative differences between the base Gemma-2B model and Gemma-2B-Tele. These examples include telecommunications-related questions from Tele-Eval (Question), and simple string prompts (Prompt).
\medskip
\medskip
\begin{tcolorbox}[colframe=niceblue!75!black, colback=blue!5!white, sharp corners=south, title=Example 1]
\textbf{Question:} What is the objective of user association in wireless communications?

\textbf{Ground Truth:} The objective of user association is to manage the assignment of users to base stations or cells, taking into account factors such as quality of service (QoS) requirements, energy efficiency, and network resources.

\textbf{Gemma-2B:} The objective of user association is to reduce the number of access points in the network.

\textbf{Gemma-2B-Tele:}  The objective of user association is to select the best radio resource for a user.

\textbf{Observation:} This example demonstrates how the model becomes more knowledgeable about specific telecommunications concepts and is able to provide more accurate answers compared to the original base model.
\end{tcolorbox}
\medskip
\medskip
\bigskip
\medskip
\medskip
\medskip
\medskip
\medskip
\begin{tcolorbox}[colframe=niceblue!75!black, colback=blue!5!white, sharp corners=south, title=Example 2]
\textbf{Question:} What does MDS stand for in the context of array codes?

\textbf{Ground Truth:} MDS stands for Maximum Distance Separable.

\textbf{Gemma-2B:}  MDS stands for Minimum Distance Decoding.

\textbf{Gemma-2B-Tele:} MDS stands for Maximum Distance Separable.

\textbf{Observation:} This example highlights the model's proficiency with telecommunications-related acronyms compared to the original model.
\end{tcolorbox}


\clearpage

\begin{tcolorbox}[colframe=niceblue!75!black, colback=blue!5!white, sharp corners=south, title=Example 3]
\textbf{Question:} What is the stopping condition used in the implementation of Wirtinger gradient descent for minimizing the objective function?

\textbf{Ground Truth:} The stopping condition used in the implementation of Wirtinger gradient descent for minimizing the objective function is when the iterates converge to a critical point.

\textbf{Gemma-2B:} The stopping condition is the condition that the gradient of the objective function is less than a certain threshold.
The gradient of the objective function is calculated using the Wirtinger gradient descent algorithm. The algorithm is a variant of the gradient descent algorithm that is used to minimize the objective function.

\textbf{Gemma-2B-Tele:} The stopping condition used in the implementation of Wirtinger gradient descent for minimizing the objective function is the following:
$$
\begin{array}{l}
\frac{\partial f(x)}{\partial x} = 0\\
\frac{\partial^2 f(x)}{\partial x^2} < 0
\end{array}
$$
where $f(x)$ is the objective function.

\textbf{Observation:} This example demonstrates that the trained model can create entire answers in runnable LaTeX equations instead of relying solely on natural language.
\end{tcolorbox}

\begin{tcolorbox}[colframe=niceblue!75!black, colback=blue!5!white, sharp corners=south, title=Example 4]
\textbf{Prompt:} Radio

\textbf{Gemma-2B:} active decay is first order in the decaying isotope. For example, strontium-90 contained in fallout from nuclear explosions decays to yttrium-90 and a beta particle.

\textbf{Gemma-2B-Tele:} -frequency identification, also known as wireless tag or transponder, is technology intended for automatic identification without physical contact or line-of-sight between tags and the readers (transceivers). RFIDs may use wireless communication protocols, such as Bluetooth, and employ modulation and encoding techniques such as frequency-shift or amplitude modulation.

\textbf{Observation:} This example demonstrates how the LLM's knowledge becomes tailored to the telecommunications domain, as evidenced by the word `radio' being associated with radiofrequency rather than radioactivity.
\end{tcolorbox}

\section{Literature Tasks Preparation}
\label{appendix:datasetprep}
We create instruction datasets for literature tasks using our collected Tele-Data. The content of each paper generates instructions for title generation, abstract completion, and introduction-to-abstract tasks. Using citation information from Tele-Data, each edge creates a positive citation link prediction instruction. For balance, we replace the target paper in each edge with a random paper to create negative citation link prediction instructions, maintaining a 1:1 ratio. For every citation sentence associated with an edge, we develop an instruction that uses the titles and abstracts of both connected nodes to generate the citing sentence. For citation recommendation tasks, we randomly sample 10 negative nodes per edge to create a candidate set, requiring the model to identify the correct positive candidate.

During the fine-tuning process, we adopt the Q-LoRa approach \cite{dettmers2023qloraefficientfinetuningquantized}, initializing the model in 8-bit precision and configuring the LoRa rank to 8. The LoRa layers are applied specifically to the $\mathbf{Q}$, $\mathbf{K}$, $\mathbf{V}$, and $\mathbf{O}$ matrices. To optimize training efficiency, we use a batch size of 8 along with a gradient accumulation step of 2. The learning rate is set to 0.0002, and optimization is performed using AdamW \cite{loshchilov2019decoupledweightdecayregularization}. Furthermore, we assign a LoRa scaling hyperparameter $lora\_alpha$ of 32 and incorporate a dropout rate of 0.05 during fine-tuning.

\end{document}